\documentclass[a4paper]{article}

\title{Reeb Complements for Exploring Inclusions Between Isosurfaces From Two Scalar Fields}

\usepackage{graphicx}

\graphicspath{{images/}}

\usepackage{algorithm}
\usepackage{algpseudocode} 
\usepackage{tikz-cd}
\usepackage{ulem}
\usepackage{hyperref}
\usepackage{xcolor}
\usepackage{amsmath}
\usepackage{amsthm}
\usepackage{amsfonts}
\usepackage{cleveref}

\newtheorem{definition}{Definition}

\newcommand{\commentinline}[2][_]{\textcolor{red}%
{%
\ifx#1_[
\else[#1: \fi
#2]}
}

\author{Akito Fujii \thanks{Kyushu University, Japan. Email: fujii.akito.834@s.kyushu-u.ac.jp} %
    \and Osamu Saeki \thanks{Kyushu University, Japan. \url{https://imi.kyushu-u.ac.jp/\string~saeki/}\\ Email: saeki@imi.kyushu-u.ac.jp} %
    \and Daisuke Sakurai \thanks{Kyushu University, Japan. Email: d.sakurai@ieee.org} %
}

\begin{document}

\maketitle

\begin{abstract}
This article proposes to integrate two Reeb graphs with the information of their isosurfaces' inclusion relation.
As computing power evolves, there arise numerical data that have small-scale physics inside larger ones – for example, small clouds in a simulation can be contained inside an atmospheric layer, which is further contained in an enormous hurricane.
Extracting such inclusions between isosurfaces is a challenge for isosurfacing: the user would have to explore the vast combinations of isosurfaces $(f_1^{-1}(l_1), f_2^{-1}(l_2))$ from scalar fields $f_i: M(n) \to \mathbb{R}$, $i = 1, 2$, where $M$ is an $n$-dimensional domain manifold and $f_i$ are physical quantities, to find inclusion of one isosurface within another.
For this, we propose the \textit{Reeb complement}, a topological space that integrates two Reeb graphs with the inclusion relation.
The Reeb complement has a natural partition that classifies equivalent containment of isosurfaces.
This is a handy characteristic that lets the Reeb complement serve as an overview of the inclusion relationship in the data.
We also propose level-of-detail control of the inclusions through simplification of the Reeb complement.
We demonstrate that the relationship of two independent scalar fields can be extracted by taking the product of Reeb graphs (which we call the \textit{Reeb product}) and by then subtracting the projection of the Reeb space, which opens up a new possibility for feature analysis.
\end{abstract}

\section{Introduction}
\label{sec:introduction}


The nature is abundant with multi-scale physics, where phenomena of different scales interplay with each other \cite{klein2001asymptotic}. 
For example, a large-scale vortex like a hurricane contains a range of smaller-scale vortices such as tornadoes \cite{wurman2018role}.
Such inclusion relationships of phenomena, albeit being hard to simulate due to the complexity of numerical computation, can yet be found in simulation results as geometric features, as Bürger et al.~reported \cite{burger2012vortices}.
For investigating inclusions of phenomena, one potential approach is \textit{isosurfacing}.
Indeed, this was taken in the report.
When conducting isosurfacing, the user extracts and visualizes the preimage $f^{-1}_i(l_i)$ of some scalar field $f: M(n) \to \mathbb{R}$ and an \textit{isovalue} $l_i \in \mathbb{R}$.
Here, $M(n)$ is a manifold of $n$ dimensions.
The term \textit{isosurface component} (or \textit{contour}) refers to a connected component of the preimage.
In practice, isosurfacing is done for multiple scalar fields, each corresponding to different phenomena of interest \cite{Fuchs:2009}.
Even in case of Bürger et al., who investigated vortices, they extracted isosurfaces from multiple scalar fields that were sub-sampled.
This was done to extract isosurfaces from different physical scales of interest.
Generally speaking, inclusion relationships between two isosurfaces can be computed with a polygon-in-polygon algorithm.
At the first glance, it looks like the user would be satisfied.

However, when doing isosurfacing, the user has to choose the value for each parameter $l_i$.
Even for the bi-field case, where $\{f_i\}_i = \{f_1, f_2\}$, the user would desire to understand the inclusion relationship between two isosurfaces with different choice of $(l_1, l_2)$ from $\mathbb{R}^2$.
In practice, each $l_i$ can be a floating point number, and thus results in virtually infinite number of choices, where a slight change in either $l_i$ value might give a drastic difference in the extracted isosurfaces and their inclusion relation.
Such a challenge is, indeed, described also in the aforementioned report.
One step forward into solving this problem can be taken by computing the \textit{Reeb graph} \cite{bajaj1997contour,parsa2012deterministic, reeb1946singular} well-known for aiding isosurfacing \cite{bajaj1997contour,carr2010flexible,tierny2017topological}.
(If the domain is simply connected, the Reeb graph is often called the \textit{contour tree} \cite{bajaj1997contour}.)
The Reeb graph is a topological space constructed by identifying points in the same isosurface as a single point, as will be explained in \Cref{sec:preliminary}.
A single Reeb graph serves as an overview of data for a single $f_i$, describing how isosurfaces in $f_i$ get morphed as $l_i$ varies in $\mathbb{R}$ \cite{bajaj1997contour,carr2010flexible}.
The Reeb graph is not a complete solution to our challenge, since it does not tell us the inclusion relationship.
Our problem, then, is to integrate the information of inclusion with that of the aforementioned overview solution, i.e.~the Reeb graphs.

\begin{figure}[t]
    \includegraphics[keepaspectratio, width=\linewidth]{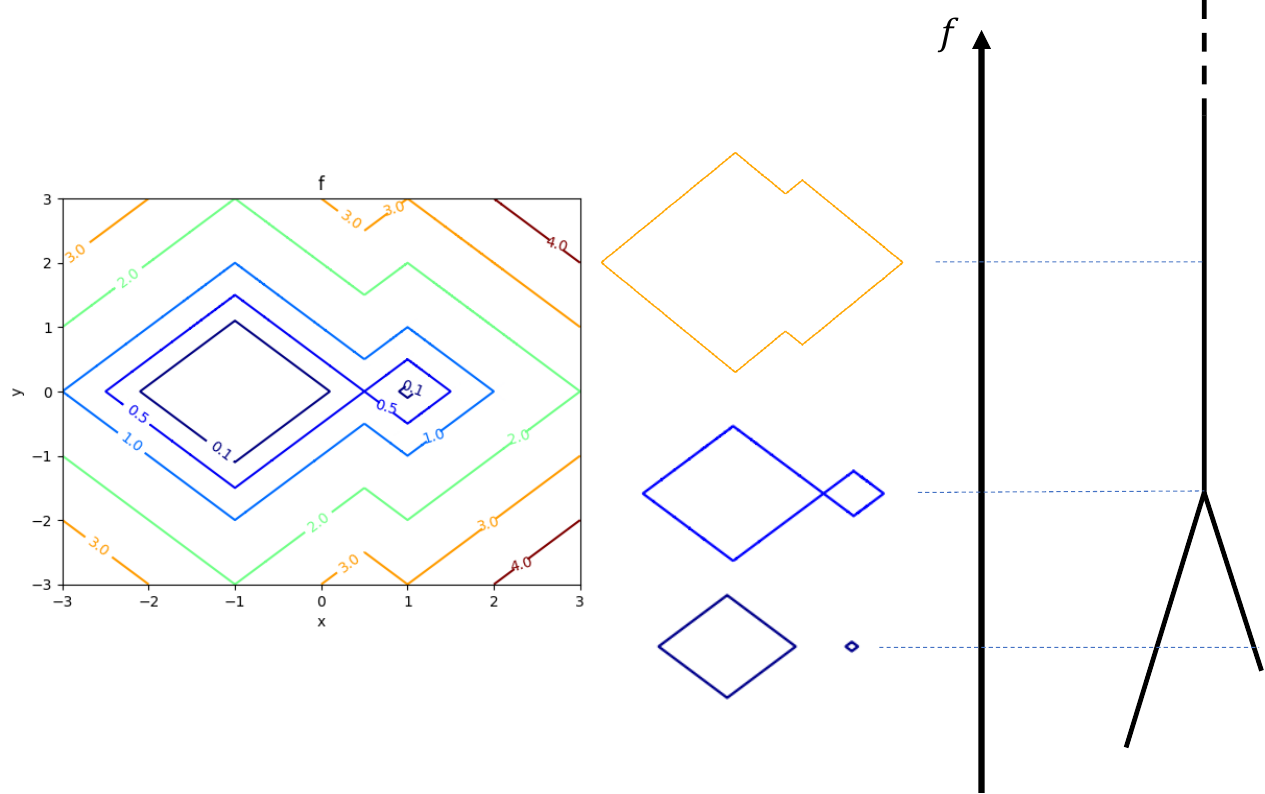}
    \centering
    \caption{
        \label{f}
        \label{fig:example_RG}
        The Reeb graph summarizes the landscape of a height field $f: M \to \mathbb{R}$. Each contour, i.e.~a connected component of a preimage of $f$, becomes a point in this 1-dimensional topological space illustrated in this figure.
    }
\end{figure}


We thus propose another dedicated topological space, which we call the \textit{Reeb complement}.
The Reeb complement integrates the Reeb graphs with the possible inclusion relations by associating each point (i.e.~isosurface) in one Reeb graph with a point in the other graph, corresponding to an isosurface disjoint from the original one (see \Cref{sec:reebcomplement}).
The Reeb complement summarizes inclusions in data by partitioning points with similar inclusions in the same partition.

Thanks to this, the Reeb complement can serve as an overview that navigates the exploration of isosurface inclusions in the data, much like the Reeb graph is an overview that does so for the exploration of isosurfaces.
We see two ways to utilize the Reeb complement: either it can be used internally to find interesting isosurface intersections or it can be directly shown to the user to show a summary of the isosurface inclusions \cite{carr2010flexible}.
These applications, however, require further studies, and are thus out of the scope of this article.

As the name suggests, the Reeb complement can be seen as the complement of the Reeb space (see \Cref{sec:reebcomplement}).
We also propose level-of-detail control of the inclusions through simplification of the Reeb complement.

\subsection{Related Work}\label{sec:related}

The Reeb graph has been studied extensively \cite{parsa2012deterministic, pascucci2007robust, bajaj1997contour,carr2003computing, doraiswamy2009efficient}. In this section, we will see that existing work either did not take into account two Reeb graphs or did not extract inclusion relations.

Takahashi et al.~\cite{takahashi2004topological} augmented a single contour tree with the inclusion relations between isosurfaces from a \emph{single} scalar field $f: \mathbb{R}^3 \to \mathbb{R}$, while we consider two of them instead.
Their algorithm identified the change to the inclusion of an isosurface in another by analyzing the saddle point when these merge into one.
A similar approach was taken by Pascucci et al.~\cite{Pascucci2004} to decide the homology of isosurfaces, although this did not identify the inclusion.
For $f: \mathbb{R}^2 \to \mathbb{R}$, inclusion relations between isosurfaces were studied by Guibert \cite{guilbert2013multi}.

Schneider et al.~\cite{schneider2008interactive} proposed to combine two Reeb graphs of two scalar fields.
They extracted a number of isosurfaces from two fields and analyzed the overlap between the isosurfaces' volume inside.
While such overlap information can correlate with information of inclusion, they do not match.

Another difference is that Schneider et al.~pre-selected a finite number of isosurface pairs to analyze, while our Reeb complement allows all pairs.
The preselection by Schneider et al.~was a deliberate choice to provide an overview of isosurface pairs that were deemed similar in their formulation.
In our case, we allow all pairs since we can then allow the user of isosurfacing to investigate any of them without losing the context (i.e.~the summary of isosurface inclusions).
Of course, the engineer who implements the Reeb complement is free to offer the same pre-selection following Schneider et al., and those pairs can be still located in the Reeb complement.

Finally, the Reeb space \cite{edelsbrunner2008reeb,tierny2016jacobi} is yet another topological space, whose points correspond to the intersections between isosurfaces from different scalar fields.
It has approximations such as the mapper \cite{munch2015convergence} and joint contour net \cite{carr2013joint}.
The Reeb space by itself does not tell us the inclusion information we are after, yet is a building block of our formulation (see \Cref{sec:reebcomplement}).

\section{Preliminary} \label{sec:preliminary}

Let $f$ be a map from the domain $M$ to $\mathbb{R}^n$.
The domain can be a manifold or its triangulation: in the former case $f$ is smooth, and in the latter, it is piece-wise linear (PL), and $M$ may have boundary.
We define the equivalence relation $\sim$ on $M$ in such a way that $x \sim y$ for $x, y \in M$ whenever $x$ and $y$ belong to the same isosurface.
The \textit{Reeb space} $W_f$ is then defined as the quotient space $M / \sim$.
We have the quotient map $q_f : M \rightarrow W_f$ and the map $\bar{f} : W_f \rightarrow \mathbb{R}$ such that $f = \bar{f} \circ q_f$. Note that the preimage of each point in $W_f$ by $q_f$ is a connected component of a preimage of $f$.
If the range of $f$ is $\mathbb{R}$, $f$ is a scalar field, and the Reeb space is called the \textit{Reeb graph} and is denoted by $G_{f}$.
In practice, $G_f$ can be assumed to have the topology of a graph \cite{saeki2022reeb}.

From now on, we assume a generic scalar field $f$.
For a smooth $f$, this means that $f$ is a Morse function \cite{cole2003loops, milnor1969morse}.
For a PL $f$, $f$ is perturbed with the simulation of simplicity \cite{edelsbrunner1990simulation}.
This assumption guarantees that critical points $C \subset M$ of $f$ are isolated in the domain and so are its values $f(C)$ in the range.
In both categories, such generic scalar fields can approximate all smooth or PL maps of $M$ into $\mathbb{R}$ \cite{edelsbrunner2022computational}. 

When we map critical points into the Reeb graph using the quotient map, they appear as vertices of degree different from $2$, with the exception of saddle points that do not correspond to splitting of an isosurface into multiple ones, nor to merging of multiple contours into a single one.
Inside $M$, only at critical points can the topological shape of an isosurface change by changing the isovalue.

For example, let us consider $\mathbb{R}^2$ as the domain and the smooth map $f : \mathbb{R}^2 \rightarrow \mathbb{R}$ defined as follows:

\begin{equation}
    f = \left \{
    \begin{array}{l}
    |x-1| + |y|~ (x \geq 1/2) \\
    |x+1| + |y| - 1~ (x < 1/2).
    \end{array}
    \right.
\end{equation}

This $f$ is illustrated in Figure \ref{f}.
As we raise the isovalue from the bottom of the range, an isosurface appears when $f(x, y) = -1$, which is the global minimum value.
Another isosurface appears at $f(x, y) = 0$, which is a local minimum value.
Each of them grows while $f(x, y) < 1/2$ and then they merge when $f(x, y) = 1/2$.
After that, they continue to grow with no topological change.
Thus, the Reeb graph of this $f$ is as in Figure \ref{fig:example_RG}.

In isosurfacing, it is common to change the isovalue $l_i$ to tune the visualization result.
This is done to find a reasonable geometric representation $f_i^{-1}(l_i)$ of some object of interest.

When conducting topological analysis for isosurfacing, this similarity can be conceptualized as an equivalence relation.
For convenience, let $c_i \subset f^{-1}(l_i)~(i \in \{1,2\})$ denote two isosurfaces.
These two isosurfaces are equivalent if they can be morphed into each other by continuously changing the isovalue $l$ of the isosurface $c$ from $l_1$ to $l_2$ without undergoing certain topological changes of the shape of isosurface. 

\begin{figure}[htbp]
    \centering
    \includegraphics[width=0.3\linewidth]{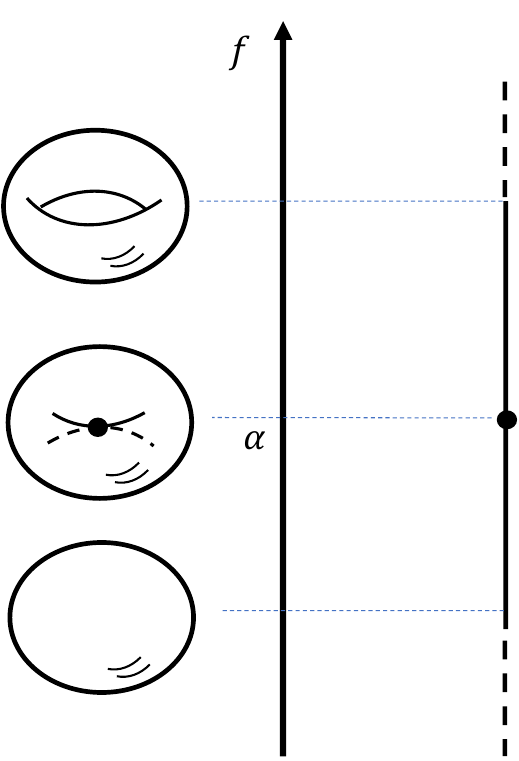}
    \caption{Example where the contour topologically changes but the number of the contours stays to be one.}
    \label{fig:topo_change_cnt}
\end{figure}

These ``certain'' changes are, in a dogmatic form of topological analysis, all changes in the topological shape of $c$. 
In practice, however, some changes are often ignored.
These ignored are, for example, changes that do not split $c$ into two, like a change in genus (see \Cref{fig:topo_change_cnt}).
So are changes at the boundary of the domain, like holes being made on the isosurface.

It is customary to mark these changes in the Reeb graph as its additional vertices that divide some Reeb graph edges.
(While resulting variations of a Reeb graph is sometimes called an augmented Reeb graph, it is rare to find specific terms to differentiate the variations. We thus simply treat these different Reeb graphs as the Reeb graph equipped with different partitioning into its edges by adding vertices at the topological changes of interest.)
We can use the edges of the Reeb graph to define the equivalence between isosurfaces:
\begin{definition} \label{def:equivalentcontours}
    Isosurfaces $c_1 \subset f^{-1}(l_1)$ and $c_2 \subset f^{-1}(l_2)$ %
    are said to be \textbf{equivalent} if $q_f$ maps $c_1$ and $c_2$ onto the same edge of the Reeb graph.
\end{definition}
This equivalence gives the equivalence classes that correspond to features such as the representative contour \cite{takahashi2004topological} and flexible isosurface \cite{carr2010flexible}.







\section{Isosurface Inclusions} \label{sec:partition}

\begin{figure}
  \centering
      \begin{tabular}{cccc}
         \includegraphics[width=0.3\textwidth]{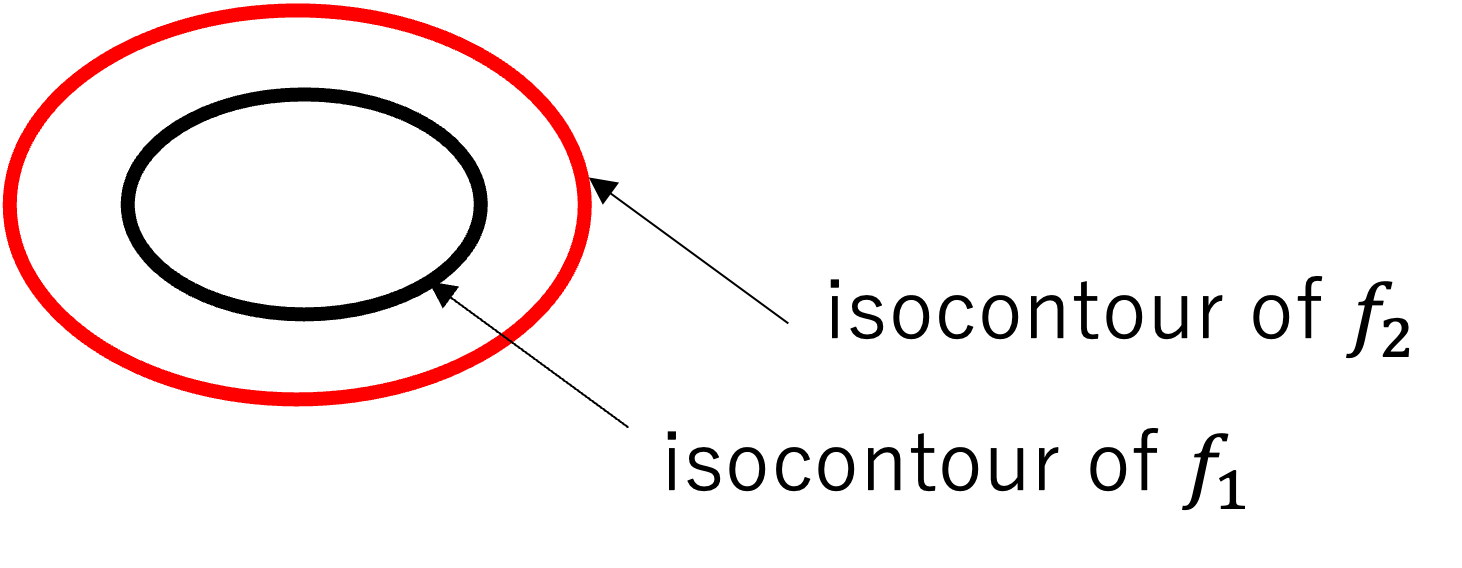} &
         \includegraphics[width=0.3\textwidth]{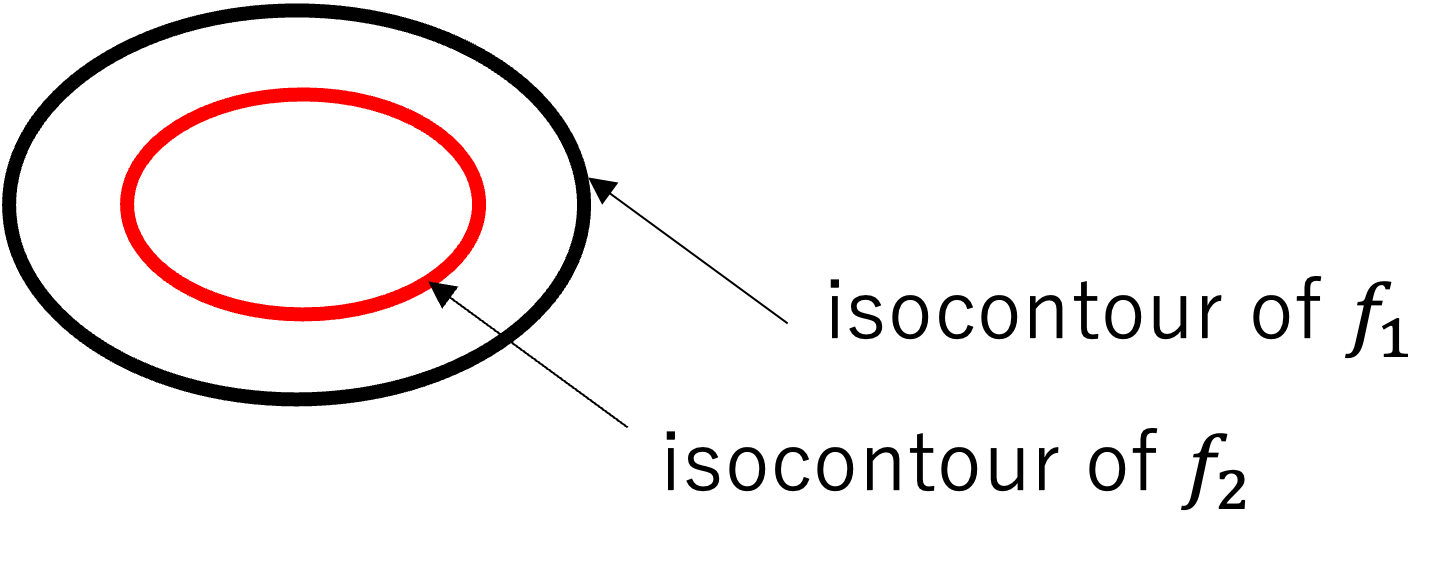} &
         \includegraphics[width=0.3\textwidth]{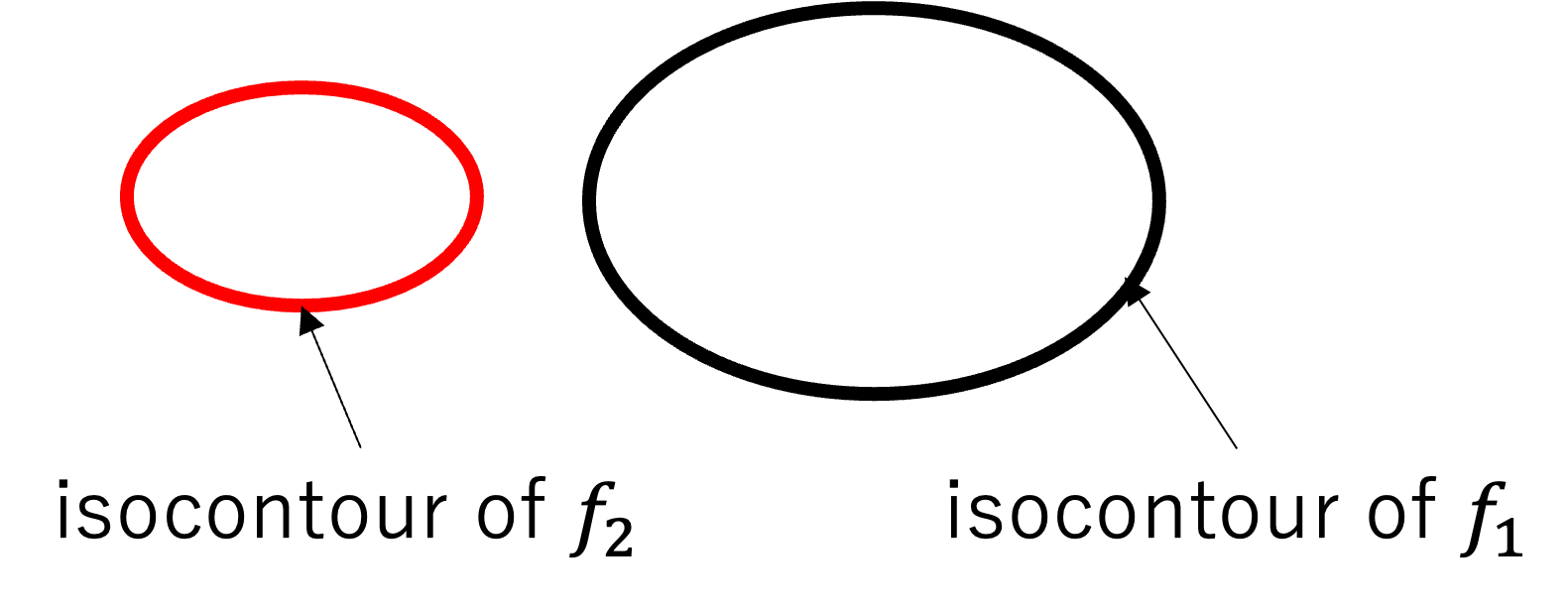} \\
         (a) & (b) & (c)
      \end{tabular}
  \caption{Two isosurfaces without intersection, in a Euclidean space. }
  \label{fig:inclusion}
\end{figure}

Imagine a pair of isosurfaces from $f_1$ and $f_2$, both defined in a $d$-dimensional manifold (or its triangulation), permitting boundary.
As such a pair exhibiting inclusion, we consider only the cases where the pair does not intersect with each other.
In the real world, even for intersecting isosurfaces, the user may perceive inclusion.
For example, the part of the isosurface $c_1 \subset f_1^{-1}(l_1) $ outside $c_2 \subset f_2^{-1}(l_2)$ can be tiny, compared to the part inside.
Treating such a complicated pair as exhibiting inclusion is left as future work.

As illustrated in \Cref{fig:inclusion}, we can decide the inside and outside of an isosurface under a common situation in isosurfacing – namely, an isosurface is a compact connected manifold (or triangulation) lying in a Euclidean space.
In this case, we have a clear notion of inclusion: an isosurface of dimension $d-1$ includes other isosurfaces that lie inside.
\textit{None includes} can also be the answer (\Cref{fig:inclusion} (c)).

\begin{figure}
  \centering
      \begin{tabular}{ccccc}
         \includegraphics[width=0.2\textwidth]{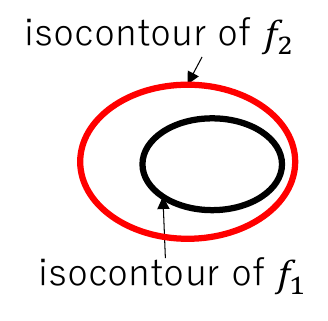} &
         \includegraphics[width=0.12\textwidth]{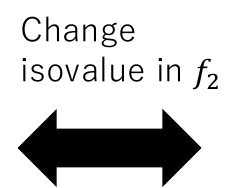} &
         \includegraphics[width=0.2\textwidth]{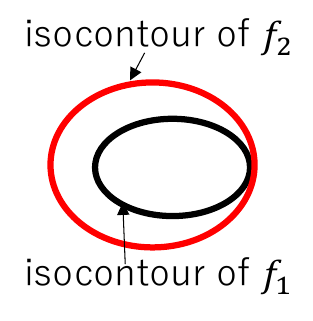} &
         \includegraphics[width=0.12\textwidth]{arrow_change_isovalue.pdf} &
         \includegraphics[width=0.2\textwidth]{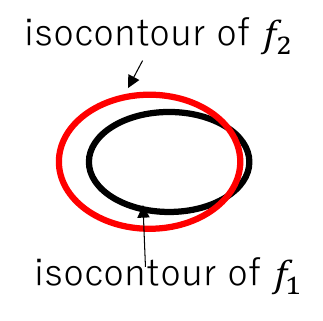} \\
         (a) & & (b) & & (c)
      \end{tabular}
  \caption{Topological change of inclusion relationship.}
  \label{fig:change_relation}
\end{figure}


We show an example of a track of topological change of inclusion relationship in Figure \ref{fig:change_relation}.
In Figure \ref{fig:change_relation}, (a) is an inclusion relationship, while (c) is not.
As indicated, the inclusion relationship breaks down when two isosurfaces touch each other at (b).
It also stands that inclusion relationships do not change as long as two isosurfaces do not interesect.

Summarizing the above, we decide when a non-intersecting isosurface pair is similar to another pair.
We, again, express this as an equivalence relation, much like we did for isosurfaces in \Cref{def:equivalentcontours}:
\begin{definition} \label{def:equivalent}
    Two pairs of isosurfaces $(c_1^1, c_2^1), (c_1^2,c_2^2)$ have an \textbf{equivalent inclusion} if $c_i^1$ and $c_i^2$ are equivalent for $i = 1,2$ and $c_i^1$ can be morphed into $c_i^2$ (following an edge of the Reeb graph of $f_i$, $i = 1, 2$) by following a path in the range of $f= (f_1,f_2)$ while avoiding intersection between the two isosurfaces from $f_1$ and $f_2$.
\end{definition}

Notice that equivalent pairs have the same inclusion relationship from the three possibilities found in \Cref{fig:inclusion}.

\section{Reeb Complement} \label{sec:reebcomplement}
\label{sec:defComplement}


Now that we formulated the similarity of inclusions in terms of equivalence, we proceed to define the Reeb complement, which encodes all the equivalent isosurface inclusions in the data.
Since we consider inclusion between a pair of isosurfaces without an intersection, we can ignore pairs of isosurfaces \emph{with} intersection.
The intersections of isosurfaces are given by the Reeb space, and the set of pairs of isosurfaces of different fields can be formulated as the product space of the two Reeb graphs:

\begin{definition}
    The product space $G_{f_1} \times G_{f_2}$ is called the Reeb product.
\end{definition}

We would like to know which pairs of isosurfaces have intersections.
For this purpose, we define the map $\phi$ from the Reeb space to the Reeb product in the following manner.

Let $f = (f_1, f_2)$ be a map from the domain $M$ to $\mathbb{R}^2$.
Again, $f$ can be either smooth or PL.
We do not constrain $f$ to be generic, so that we can define the Reeb complement for an arbitrary $f$.
We denote the Reeb graph of the scalar field $f_i$ as $G_{f_i}$, $i = 1, 2$, and the Reeb space of the field $f$ as $W_f$.
We consider the map $\phi$ from the Reeb space $W_f$ to the Reeb product $G_{f_1} \times G_{f_2}$ defined as follows.
We have the quotient map $q_f : M \rightarrow W_f$ and the map $\bar{f} : W_f \rightarrow \mathbb{R}^2$ such that $f = \bar{f} \circ q_f$.
Each point in the Reeb space is an intersection component between the $f_1$'s isosurface $c_{f_1}$ and $f_2$'s isosurface $c_{f_2}$. In practical terms, $\bar{f}$ maps the point of the Reeb space to $(v_1, v_2) \in \mathbb{R}^2$, where $v_1$ is the isovalue of $c_{f_1}$ and $v_2$ is the isovalue of $c_{f_2}$.
On the other hand, a point of the Reeb graph is an isosurface.
This $\bar{f_i}$ maps the point in the Reeb graph to the value of the isosurface in $\mathbb{R}$:

\begin{center}
\begin{tikzcd}
    M \arrow[rr, "f"] \arrow[rd, "q_f"]  & & \mathbb{R}^2 & M \arrow[rr, "f"] \arrow[rd, "q_{f_i}"]  & & \mathbb{R} \\
& W_f \arrow[ru, "\bar{f}"] & & & G_{f_i} \arrow[ru, "\bar{f_i}"] &.
\end{tikzcd}
\end{center}

Let $w$ be an element of $W_f$.
Let $x$ be an arbitrary point of $q_f^{-1}(w)$.
$q_f^{-1}(w)$ is the fiber component, that is, a connected component generated by the intersection between single $f_1$'s isosurface $c_1$ and single $f_2$'s isosurface $c_2$.
$q_f^{-1}(w)$ is connected so the arbitrary element $x \in q_f^{-1}(w)$ is on $c_1$ and $c_2$.
By projecting $x$ by the map $q_{f_i}$, we can obtain the isosurface $c_i$.
Therefore, we can write the map $\phi$ as $\phi(w) = (q_{f_1}(x), q_{f_2}(x))$.
Thus, the map $\phi$ is the unique map such that $(q_{f_1}, q_{f_2}) = \phi \circ q_f$ in the diagram below:

\begin{center}
    
\begin{tikzcd}
M \arrow[rr, "{(q_{f_1}, ~q_{f_2})}"] \arrow[rd, "q_f"]  & & G_{f_1} \times G_{f_2} \\
& W_f \arrow[ru, "\phi"] &.
\end{tikzcd}

\end{center}

\begin{definition}
    The \textbf{Reeb complement} is 
 $(G_{f_1} \times G_{f_2}) \setminus \phi(W_f)$.
\end{definition}
Although we could define the Reeb complement for an arbitrary dimensional range as $\prod_i G_i \setminus \phi(W_f)$, doing so is out of the scope of the present article.

\section{Partitioning the Reeb Complement}

We now partition the Reeb complement, so that the resulting partition cells correspond to the sets of equivalent inclusions of isosurfaces as defined in \Cref{def:equivalent}.
We first partition the Reeb complement using the rectangles consisting of all pairs of edges in $G_{f_1} \times G_{f_2}$.
Inside each rectangle, the isosurfaces $c_i$ of $f_i$ are all equivalent for the field $i$ (recall \Cref{def:equivalentcontours}).
On the contrary, these isosurfaces are unequivalent to any other.

Inside this rectangle, we have the region corresponding to $\phi(W_f)$.
Mapped to these regions, which are outside the Reeb complement, are the intersecting pairs of isosurfaces.
Summarizing all the above, each connected component of the Reeb complement in each rectangle in fact corresponds to one equivalence class of isosurface inclusions.

As explained in \Cref{sec:preliminary}, we allow the edges of the Reeb graph to reflect the features of interest, such as whether topological changes in isosurface shapes at the boundary are to be ignored.





\section{Examples}

Before we go to the computation, we see some examples of the Reeb complement in order to familiarize ourselves with the concept.


\subsection{Isosurfaces in Another One}

Here, we investigate an example of a Reeb complement, where the level set of $f_2$ can have two isosurfaces:
\begin{equation}
    f_1 = |x| + |y|,~
    f_2 = \left \{
    \begin{array}{l}
    |x-1| + |y|~ (x \geq 0) \\
    |x+1| + |y|~ (x < 0).
    \end{array}
    \right.
\end{equation}

\begin{figure}[htbp]
    \centering
    \includegraphics[keepaspectratio, width=0.6\linewidth]{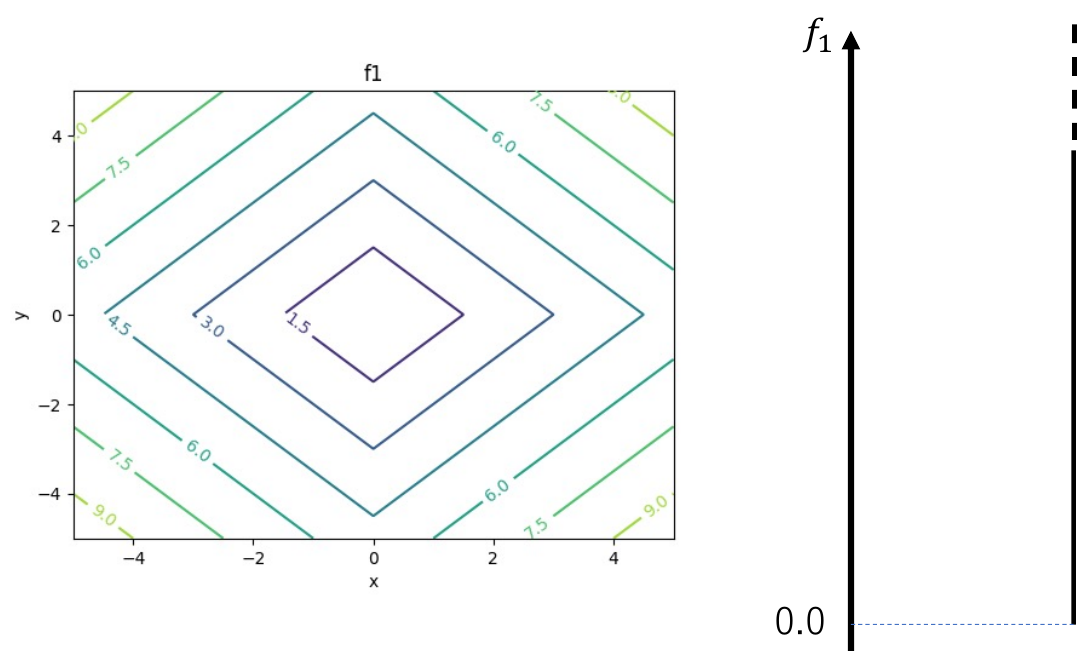}
    \caption{Contour lines of $f_1$ and the corresponding Reeb graph.}
    \label{fig:f_1}
\end{figure}

\begin{figure}[htbp]
    \centering
    \includegraphics[keepaspectratio, width=0.6\linewidth]{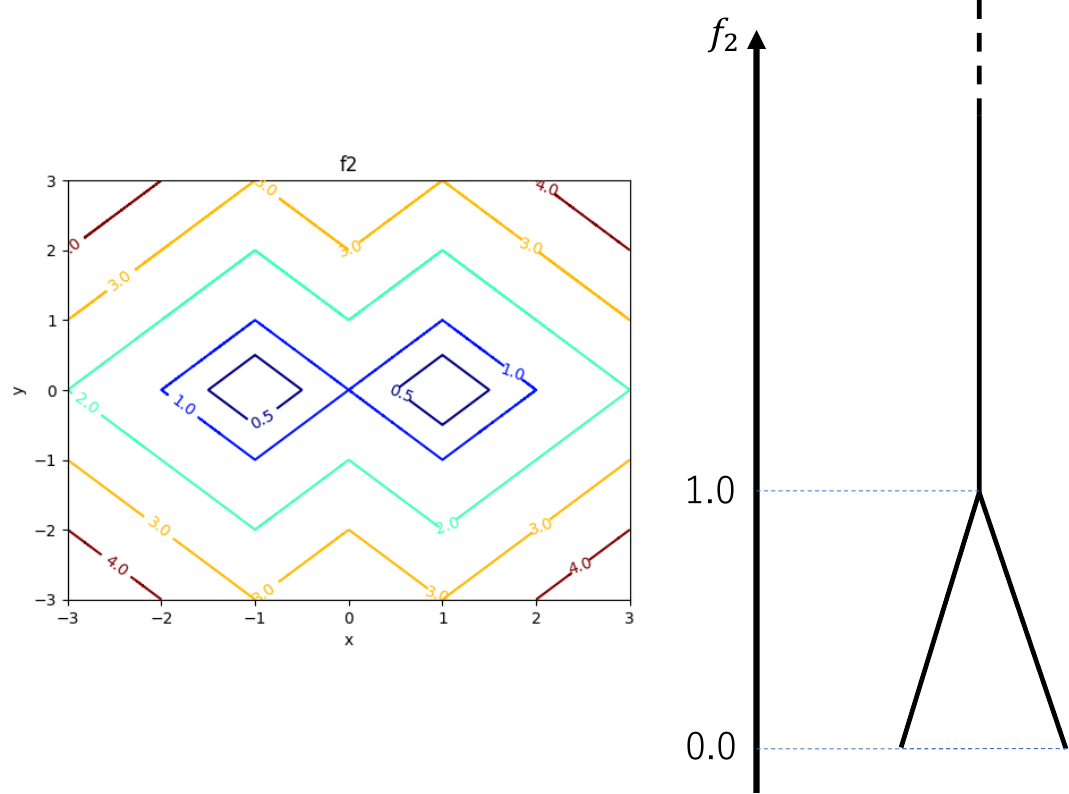}
    \caption{Contour lines of $f_2$ and Reeb graph.}
    \label{fig:f_2}
\end{figure}


\begin{figure}[htbp]
    \centering
    \includegraphics[keepaspectratio, width=0.5\linewidth]{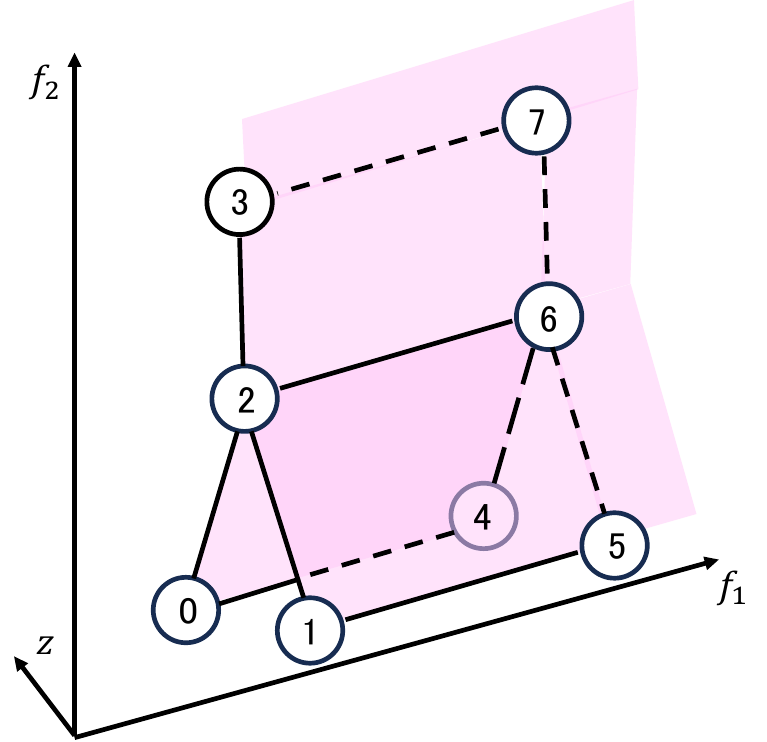}
    \caption{The Reeb product $G_{f_1} \times G_{f_2}$. We use node IDs for explaining the Reeb complement in \Cref{RC_tri1,RC_tri2,RC_parallelogram}.
    The $z$-axis is only for layouting.
    }
    \label{G_f_1_cross_G_f_2}
\end{figure}


\begin{figure}[htbp]
    \centering
    \includegraphics[keepaspectratio, width=0.7\linewidth]{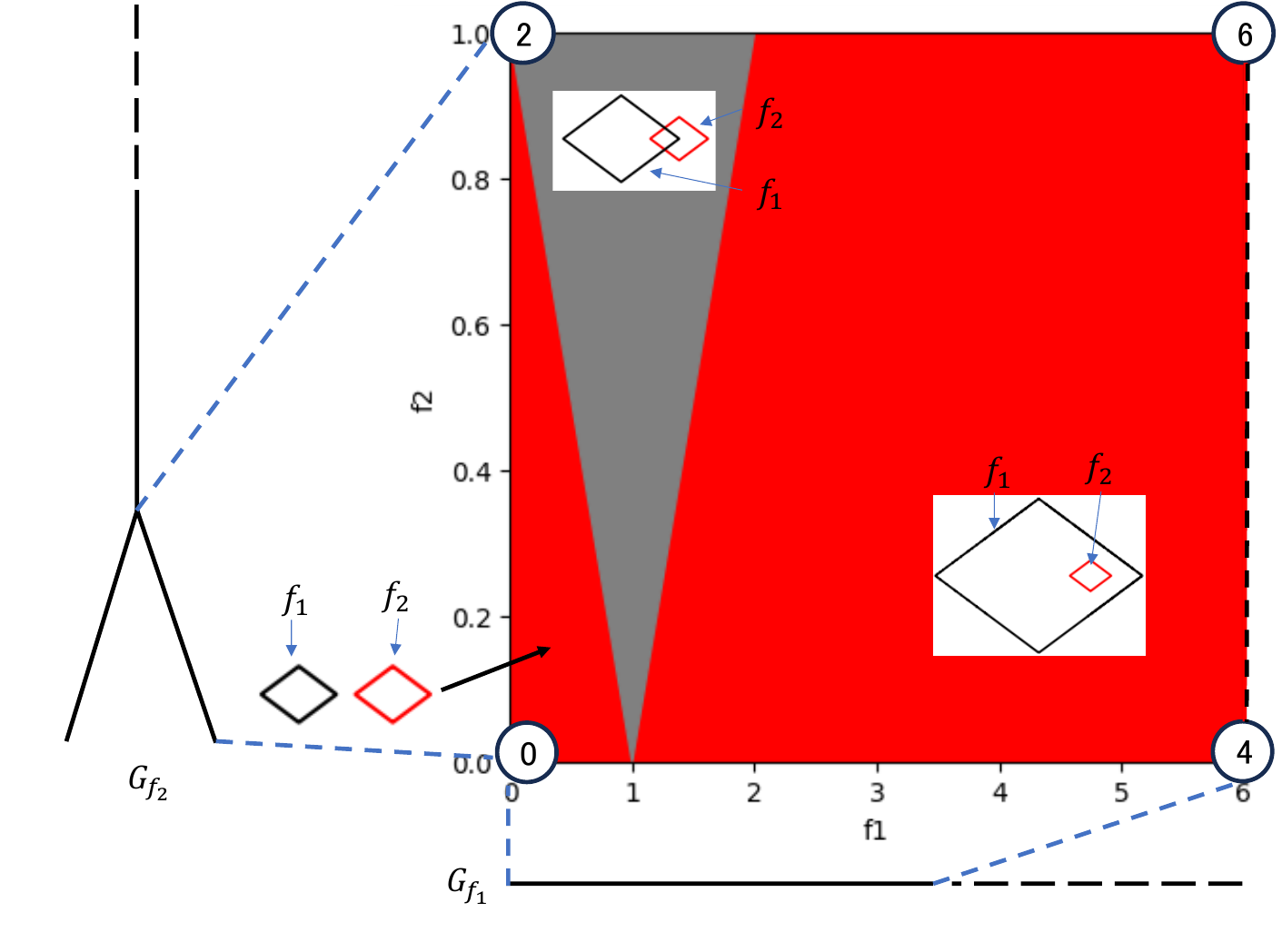}
    \caption{The rectangle consisting of vertices 0, 2, 6, 4 and the Reeb complement therein.
    The red areas correspond to the Reeb complement, and the gray area to $\phi(W_f)$. Each area corresponds to one equivalence class of inclusion.
    }
    \label{RC_tri1}
\end{figure}

\begin{figure}[htbp]
    \centering
    \includegraphics[keepaspectratio, width=0.7\linewidth]{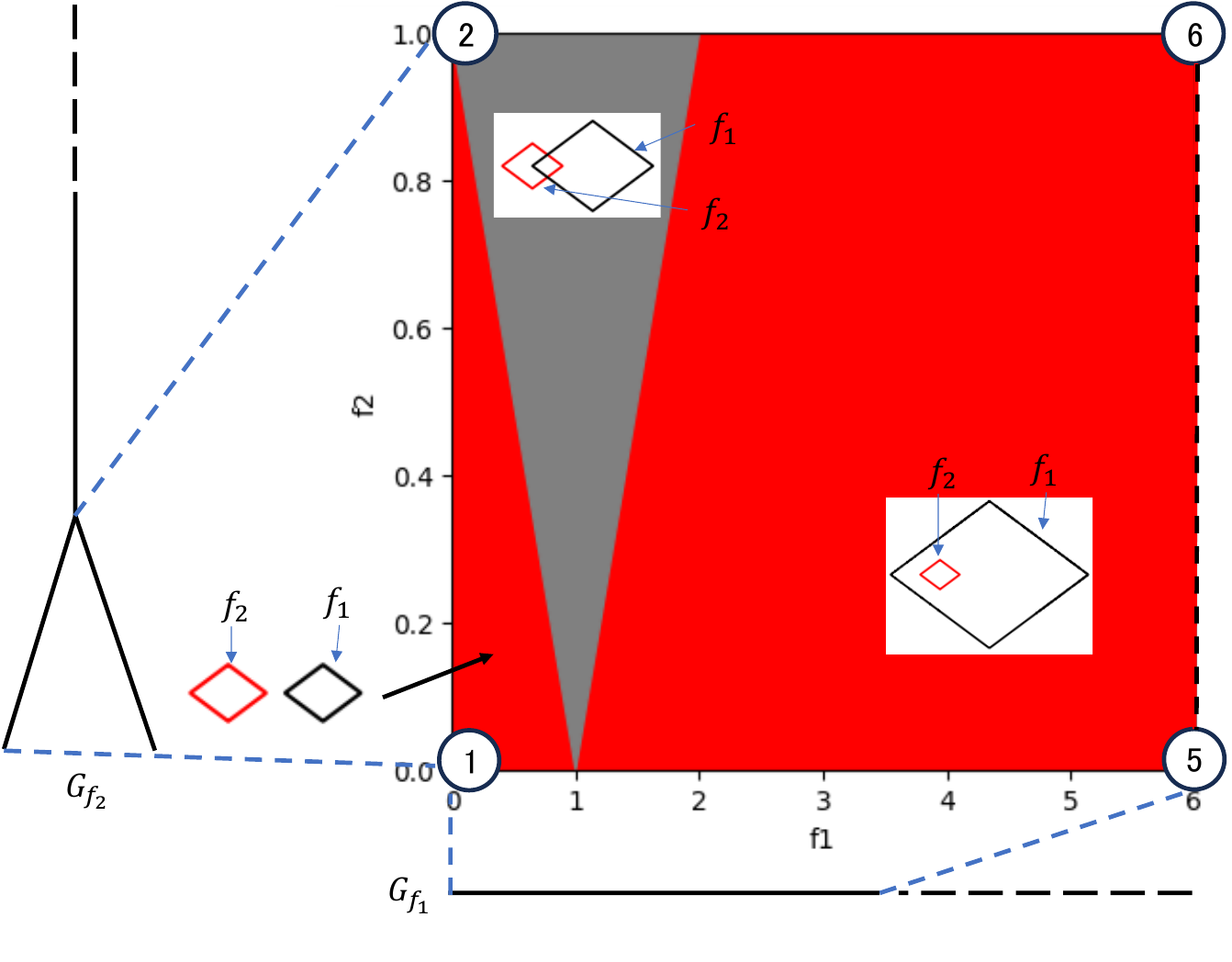}
    \caption{
        The rectangle which consists of vertices $\{ 1, 2, 6, 5 \}$ and the Reeb complement therein. The gray area corresponds to $\phi(W_f)$.
    }
    \label{RC_tri2}
\end{figure}

\begin{figure}[htbp]
    \centering
    \includegraphics[keepaspectratio, width=0.7\linewidth]{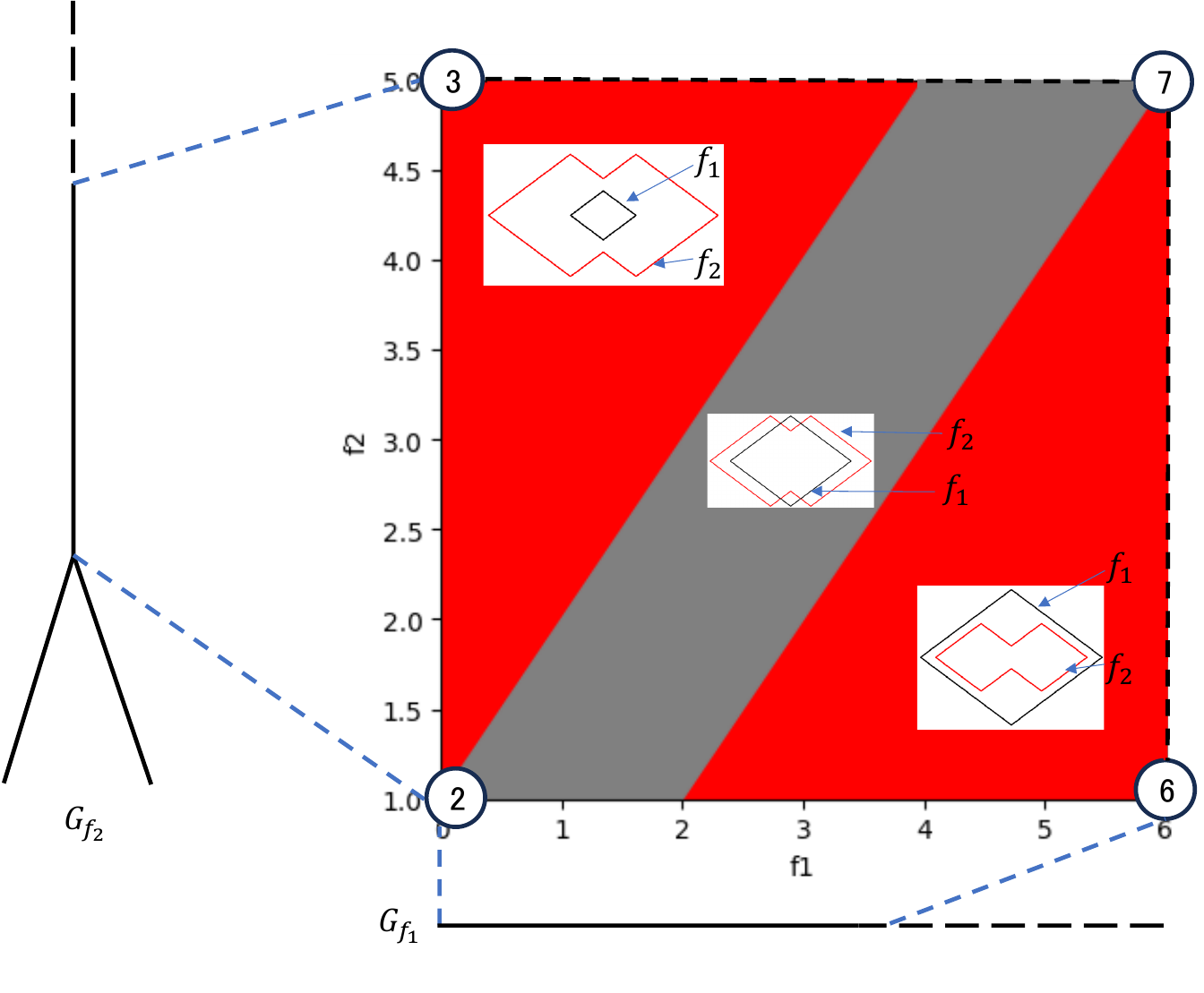}
    \caption{
        This is the rectangle which consists of vertices $\{ 2, 3, 7, 6\}$ and the Reeb complement therein. The gray area corresponds to $\phi(W_f)$.
    }
    \label{RC_parallelogram}
\end{figure}


Isosurfaces of $f_1$ and $f_2$ are illustrated in \Cref{fig:f_1,fig:f_2}, respectively.
The shape of isosurfaces of $f_1$ is always the rhombus and the number of isosurfaces is always equal to one, and so the Reeb graph $G_{f_1}$ is identical to an interval.
On the other hand, the number of isosurfaces of $f_2$ is equal to $2$ when $f_2(x, y) < 1$.
When $f_2(x, y) \geq 1$, the number of isosurfaces is equal to $1$, so $G_{f_2}$ has a figure-Y shape as in Figure \ref{fig:f_2}.
$G_{f_1} \times G_{f_2}$ is then the product of the interval and a figure-Y shape, as shown in \Cref{G_f_1_cross_G_f_2}.

We then check the Reeb space.
Let us start with the rectangle with vertices $\{ 0, 2, 6, 4 \}$ in Figure \ref{G_f_1_cross_G_f_2}.
This rectangle shows one of the two isosurfaces of $f_1$ in the corresponding range and a single isosurface of $f_2$.
Since the points of Reeb space $W_f$ correspond to intersections of isosurfaces in $f_1^{-1}$ and $f_2^{-1}$, they are the solution of $f(x, y) = (l_1, l_2)$.
We can thus map $W_f$ to the range by simply mapping the full-simplices with $f$.
In our example, this gives us the triangle area in the rectangle (\Cref{RC_tri1}).
The rectangle with vertices $\{ 1, 2, 6, 5\}$ is in the same situation (see \Cref{RC_tri2}) due to the symmetry of the two isosurfaces of $f_2^{-1}$.
Lastly, we consider the Reeb complement in the rectangle with vertices $\{ 2, 3, 7, 6\}$.
We can obtain the Reeb complement structure as in \Cref{RC_parallelogram} by tracing the intersections of contours.



\subsection{Two Identical Scalar Fields} \label{sec:pathological}

\begin{figure}
    \centering
    \includegraphics[width=0.5\linewidth]{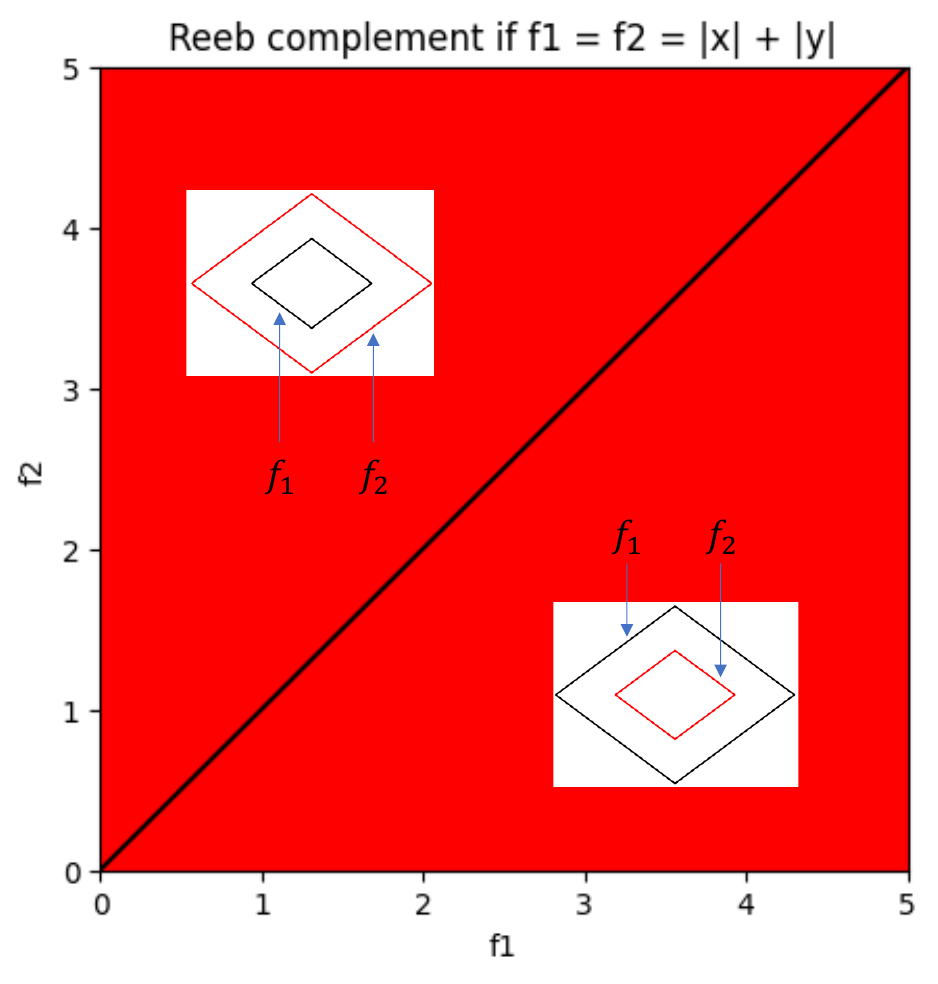}
    \caption{
        The Reeb complement corresponds to the red areas.
        The black diagonal line corresponds to $\phi(W_f)$.
    }
    \label{fig:ReebComplement2}
\end{figure}

As a pathological case, we show an example of the Reeb complement when $f_1 = f_2$.
To take an example, we assume $f_1(x,y) = f_2(x,y) = |x| + |y|$ as in Figure \ref{fig:f_1}.
The Reeb graph $G_{f_1} (= G_{f_2})$ is topologically an interval.
This gives the Reeb complement as in \Cref{fig:ReebComplement2}, divided by the diagonal line.
In the top left area, $f_2$'s isosurface includes $f_1$'s isosurface.
On the contrary, $f_1$'s isosurface includes $f_2$'s isosurface in the area in the bottom right.

\section{Algorithm}


For combinatorial computation, let us assume that $f_1, f_2: M \rightarrow \mathbb{R}$ are piecewise linear functions defined on a triangulated domain $M$, which may have boundary.
For convenience, we assume that $f_i$, $i=1,2$ underwent the \textit{Simulation of Simplicity} (SoS) \cite{edelsbrunner1990simulation}.
Without the SoS this gets complicated, as we will discuss in \Cref{sec:discussion}.

We compute the Reeb complement in the following way.

\begin{algorithm}[H]
\caption{Compute the Reeb complement for a pair of Reeb graph edges.
}
\label{alg:complement}

\begin{algorithmic}[1]

\Procedure{ComputeReebComplement}{$f$, edges $e_1,e_2$ from $G_{f_1}, G_{f_2}$}
    \State \texttt{Rectangle} $\leftarrow \bar{f_1}(e_1) \times \bar{f_2}(e_2)$ \label{line:rectangle}
    \State \texttt{Intersection $\leftarrow$ Common(Full\_simplices($e_1$), Full\_simplices($e_2$))} \label{line:intersect}
    \State \texttt{Reeb $\leftarrow$ $f($Intersection$)$} \label{line:reeb}
    \State \Return \texttt{Rectangle $\setminus$ Reeb}
\EndProcedure

\end{algorithmic}
\end{algorithm}

We assume the Reeb graphs $G_{f_1}, G_{f_2}$ to be obtained by some standard algorithm \cite{carr2003computing,pascucci2007robust}.
Each full-simplex $S$ (i.e.~$n$-simplices where $n$ is the dimension of $M$) is assigned to those edges that overlap with $q_{f_i}(S)$.
Instead of full-simplices, we could also use their $d$-faces, where $d\geq 2$.
This is because we use the image of full-simplices for $f$, which can be obtained alternatively as the union of the images of $d$-faces.
The choice would depend on the nature of triangulation $M$, provided by the user.
For the sake of explanation, the rest of the text assumes the full-simplices.

We explain the computation of the Reeb complement for a \emph{single} pair of edges in $G_{f_1}$ and $G_{f_2}$, one edge for each graph.
This is shown as \Cref{alg:complement}.
In order to use the usual boolean set operations on polygons, we map edges $e_1, e_2$ from $G_{f_1}$ and $G_{f_2}$, respectively, to the Euclidean space by $\bar{f_i}$, $i = 1, 2$, and to do geometrical computation in that space.
We start by getting the intervals $\bar{f_i}(e_i)$ in $\mathbb{R}$ in line \ref{line:rectangle}. 
We obtain the rectangle \texttt{Rectangle} as a polygon in $\mathbb{R}^2$ corresponding to the product of the intervals.

We proceed to place $\phi(W_f)$ in $e_1 \times e_2$.
The points of $M$ to be placed here reside in the full-simplices that are assigned to both $e_1$ and $e_2$.
For any such simplex $S \subset M$, the preimage of $f$ inside has a single connected component, since $f$ is PL.
Therefore, we can map this simplex with $f$ and restrict the resulting polygon within \texttt{Rectangle} to place it.
Again, due to $f$ being PL, $f(S)$ can be computed as the convex hull of the vertices of this simplex.

This is done in lines \ref{line:intersect}--\ref{line:reeb}.
There, the function \texttt{Full\_simplices} returns the full-simplices of a Reeb graph edge.
\texttt{Common} picks all the full-simplices that exist in both arguments.
If the full-simplices are given IDs, this can be done straight-forwardly by applying some set intersection algorithm.

In line \ref{line:reeb}, we map each full-simplex by $f$ as mentioned and take the union of the resulting convex hulls.
This assigns \texttt{Reeb} as this polygon the image $\phi(W_f)$ in $\mathbb{R}^2$.

Finally, to follow the definition of the Reeb complement, $G_{f_1} \times G_{f_2} \setminus \phi(W_f)$, we remove the variable \texttt{Reeb} from the rectangle.
As the boolean operation algorithm commonly uses an arrangement algorithm internally, we get the topology of the partition for free for each edge pair $(e_1, e_2)$.
For each partition, one can extract the isosurface using the marching tetrahedron algorithm, and use a point-in-polygon algorithm \cite{haines1994point} to decide which isosurface includes the other one.

If we want to get the Reeb complement for every pair of edges in $G_{f_1}$ and $G_{f_2}$, we compute the pieces of the Reeb complement in the Algorithm \ref{alg:complement} and connect them.

\section{Simplifying the Reeb Complement}

A real-world data set can contain numerous small-scale isosurfaces, which overwhelm data analysis.
These isosurfaces can be noise or small-scale features.

When dealing with the Reeb graph, it is customary to simplify the graph \cite{pascucci2007robust}.
Conceptually, this corresponds to the elimination of small-scale isosurfaces \cite{carr2010flexible}.
While it is customary to identify small-ness as the volume of the isosurface, for example, it is a parameter that can be tuned by the user, for example, using geometrical measures of isosurfaces \cite{carr2010flexible}.
The user can gradually cancel the simplification to access smaller features.

We can simplify the Reeb graph before computing the Reeb complement.
This strategy provides a clear interpretation in the context of Reeb graph-based isosurfacing.
We can, instead, compute the identical simplification by gradually simplifying the Reeb complement itself, without simplifying the input Reeb graphs.
We will describe such an algorithm later.

Aside from these two computational strategies, we have two ways to interpret the simplified Reeb graph.
In one interpretation, each isosurface assigned to the removed edges is ignored in the analysis.
In the other, the same isosurface is treated as part of another remaining isosurface \cite{carr2010flexible}.
The result is, in effect, the merging of the removed edge with another.

\begin{figure}[htbp]
    \centering
    \includegraphics[keepaspectratio, width=0.4\linewidth]{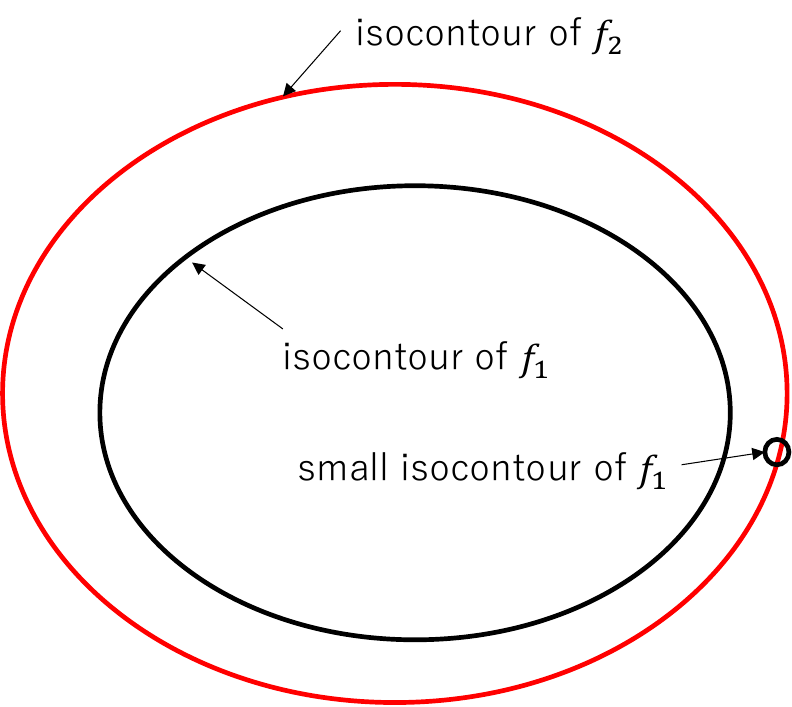}
    \caption{There is a small isosurface of $f_1$. This isosurface intersects with an isosurface of $f_2$.}
    \label{fig:small_isosurface}
\end{figure}

For example, in \Cref{fig:small_isosurface}, we merge triangles in the domain which correspond to a small isosurface of $f_1$ with the information of an edge which corresponds to a bigger isosurface of $f_1$.

The characteristics of the simplified Reeb complement if we merge contours or ignore contours are listed in Table \ref{tab:character_simp}.
In essence, if we ignore removed isosurfaces, the simplified Reeb complement would be robust against noise, since intersections due to small-scale isosurfaces are ignored in the Reeb space.
Yet, including such small-scale intersections, instead, gives a more precise Reeb complement that conveys the smallest details of the isosurface inclusions.

\begin{table}[htbp]
    \centering
    \caption{Characteristics of Reeb complement simplification.}
    \label{tab:character_simp}
    \begin{tabular}{c|c|c}
    \hline
                                 & Precision & Noise \\
    \hline
       Consider eliminated isosurfaces  & precise   & intolerant \\
    \hline
       Ignore eliminated isosurfaces & imprecise & tolerant
    \end{tabular}
\end{table}

These two variations can be realized by transferring, or not, the triangles from the removed edge to the edge preserved in the simplification of the Reeb graph, as we will see next.


\subsection{Simplification Algorithm}

Simplifying the input Reeb graph is a straightforward matter of applying a proper algorithm \cite{carr2010flexible,pascucci2007robust}.
We, hence, explain how we can directly simplify the Reeb complement.

Without loss of generality, we can assume that edges $e_1$ and $e_2$ from $G_{f_1}$ are merged and then a new edge $e_3$ is inserted in $G_{f_1}$.
Specifically, we remove $e_1$ and \emph{merge} $e_2$ into $e_3$, meaning that the full-simplices of $e_2$ are transferred to $e_3$.
If both Reeb graphs are to be simplified, we can simplify one graph at a time.

From the Reeb complement, we remove the rectangles that is a product of edge $e_1$ and another.
We, then, reconnect the rectangles as we would for the Reeb graph being simplified.

This completes the operation in case we ignore the isosurface intersections that were in the removed isosurfaces.
If we, instead, choose to reflect the intersections in the simplified Reeb complement, the following operations are done.
For each rectangle $e_1 \times e_4$, where $e_4$ is an edge of $G_{f_2}$, we take the full-simplices that were mapped to $e_1 \times e_4$ in \Cref{alg:complement} and re-map them to $\mathbb{R}^2$ with $f_1$.
We then subtract this remapped area, as it corresponds to the Reeb space mapped in the rectangle.

\section{Discussion} \label{sec:discussion}


We start by discussing how one could use the Reeb complement for data analysis in practice.
As Reeb graphs are generally not embeddable in $\mathbb{R}^2$, the Reeb complement, which is contained in their product, is also challenging to visualize in a computer screen.
In addition, the product also has numerous rectangles, which may lead to a problem on its own.
Therefore, it is challenging for the user to manually investigate the Reeb complement in its entirety.

Instead, we believe it is more straightforward to make use of the Reeb complement in case the user has visualized isosurface inclusions individually and wants to explore alternate isovalues.
In such a case, we can, for example, show the partition that contains the user-chosen pair of isosurfaces. (If two or more isosurfaces are visualized, we would need to show one for each.)
Each partition shows equivalent inclusions with different geometrical configurations of isosurfaces.
This would help the user conduct a fair analysis, in which the user is aware of possible interpretations of inclusions.
By showing nearby partitions, the user can also be notified of alternative inclusion relationships, and how the isosurface inclusions would break.
In fact, such complex inclusions were analyzed by Bürger et al.~\cite{burger2012vortices}.

By using the Reeb complement internally, the software could also suggest alternatives, picking inclusions in the same partitions and outside, equipping them with such semantic information.

In \Cref{alg:complement}, we assumed $f$ to be perturbed with the SoS.
Without this assumption, examples like the pathological case in \Cref{sec:pathological} would be rather challenging.
The Reeb space was a diagonal line, and so \Cref{alg:complement} would ignore the line and merge the two partition cells into one, generating a wrong output.

\section{Conclusion}

We proposed the Reeb complement, a topological structure that encodes inclusions between isosurfaces in two scalar fields.
The Reeb complement has a natural partition that classifies topologically equivalent isosurface inclusions.
We also proposed an algorithm to compute the Reeb complement for a pair of Reeb graph edges, as well as that for simplification, employing widely available boolean set operations of polygons.

\section*{Acknowledgement}

This work was supported in part by JSPS KAKENHI Grant Numbers JP22K18267, JP23H05437 and by WISE program (MEXT) at Kyushu University.


\bibliographystyle{IEEEtran}
\bibliography{lipics-v2021-sample-article}


\end{document}